\documentclass[review]{elsarticle}
\usepackage{lineno,hyperref}
\modulolinenumbers[5]
\usepackage{amsmath,latexsym}
\usepackage{amsfonts}
\usepackage{amssymb}
\usepackage{amssymb}
\usepackage{array} 					
\usepackage{epstopdf}
\usepackage{subfigure}
\usepackage{color}
\usepackage{setspace}
\usepackage{algorithm}
\usepackage{algpseudocode}
\usepackage[all]{xy} 				
\usepackage{tikz}
\usetikzlibrary{positioning}
\usetikzlibrary{patterns}

\usepackage{url}
\usepackage{soul}
\newtheorem{thrm}{Theorem}
\newtheorem{lemm}[thrm]{Lemma}
\newtheorem{remrk}[thrm]{Remark}

\newtheorem{obs}[thrm]{Observation}

\def\qedbox#1#2{\vbox{\hrule height.2pt
  \hbox{\vrule width.2pt height#2pt \kern#1pt \vrule width.2pt}
  \hrule height.2pt}}
\def\qed{\hfill \quad\qedbox46\newline\smallbreak}

\def\s#1{\mbox{\boldmath $#1$}}

\def\+{\!+\!}
\def\-{\!-\!}

\def\m{\!-\!}

\def\itbf#1{\textit{\textbf{#1}}}

\def\bbreak{{\bf break\ }}

\def\bto{{\bf to\ }}
\def\bdownto{{\bf downto\ }}
\def\bwhile{{\bf while\ }}

\def\band{{\bf and\ }}
\def\bor{{\bf or\ }}

\def\bbreak{{\bf break\ }}

\def\la{\leftarrow}

\def\com#1{{\bf $\triangleright$}\hspace{6pt}{\sl #1}}

\def\pref(#1,#2){$#1$ is a prefix of $#2$}
\def\suff(#1,#2){$#1$ is a suffix of $#2$}

\def\reg(#1,#2){$#2$ is $#1$-regular}
\def\notreg(#1,#2){$#2$ is not $#1$-regular}

\def\true{\tt{true}}
\def\false{\tt{false}}
\def\UPDATE\_F{\tt{UPDATE\_F}}

\def\O{\mathcal{O}}

\def\Q'{\tt{Q'}}

\newif\ifShow
\Showfalse

\algnewcommand{\LineComment}[1]{\State \(\triangleright\) \normalfont{\sl #1}}
\algtext*{EndWhile}
\algtext*{EndIf}
\algtext*{EndFor}
\algtext*{EndFor}
\algtext*{EndProcedure}

\begin{document}

\begin{frontmatter}

 \title{Practical KMP/BM Style Pattern-Matching on Indeterminate Strings}

\author[inst1]{Hossein Dehghani}
\ead{dehghh1@mcmaster.ca}
\author[inst2]{Thierry Lecroq}
\ead{thierry.lecroq@univ-rouen.fr}
\author[inst1]{Neerja Mhaskar\corref{mycorrespondingauthor}}
\cortext[mycorrespondingauthor]{Corresponding author}
\ead{pophlin@mcmaster.ca}

\author[inst1]{W.\ F.\ Smyth}
\ead{smyth@mcmaster.ca}

\address[inst1]{Algorithms Research Group,\\ Department of Computing \& Software, McMaster University, Canada}
\address[inst2]{Univ Rouen Normandie, INSA Rouen Normandie, Universit\'e Le Havre Normandie, Normandie Univ, LITIS UR 4108, F-76000 Rouen, France}

\begin{abstract}
We describe three algorithms for identifying matches of an
indeterminate pattern in an indeterminate text, both 
encoded so that indeterminate entries of up to $9$ letters
can be accommodated in a single computer word.  The two proposed algorithms, \textsc{KMP-Indet} and \textsc{BM-Indet}, extend well-known algorithms, Knuth-Morris-Pratt and and Boyer-Moore.  The
third is a simple ``brute force'' algorithm (\textsc{BF}).  All three algorithms are
fully implemented.  Recently several other such algorithms
have been proposed, including the \textsc{DBWT} algorithm~\cite{DGGL19} with a (partial) implementation.  We show that all our algorithms
execute an order-of-magnitude faster than \textsc{DBWT} on
randomly-generated strings over different alphabet sizes.
\end{abstract}

\end{frontmatter}


\section{Introduction}\label{sec:intro}
Given a fixed finite \itbf{alphabet} $\Sigma = \{\lambda_1,\lambda_2,\ldots,\lambda_{\sigma}\}$ of \itbf{size} $\sigma$, a \itbf{regular letter}, also called a \itbf{character}, is any single element of $\Sigma$, while an \itbf{indeterminate letter} is any subset of $\Sigma$ of cardinality greater than one. Let $\Sigma'$ be the set of all non-empty subsets of $\Sigma$. Then, a \itbf{regular string} $\s{x} = \s{x}[1..n]$ on $\Sigma$ is an array of regular letters of length 
$|\s{x}| = n$ drawn from $\Sigma$.
An \itbf{indeterminate} (also \itbf{degenerate})  \itbf{string} $\s{x}[1..n]$ is an array of letters drawn from $\Sigma'$, of which at least one is indeterminate. 
Whenever $\s{x}[i] \cap \s{x}[j] \neq \emptyset$, $1 \le i,j \le n$, we say that $\s{x}[i]$ \itbf{matches} $\s{x}[j]$ and write $\s{x}[i] \approx \s{x}[j]$. Two strings $\s{x}$ and $\s{y}$ \itbf{match} if and only if $|\s{x}| = |\s{y}|$ and $\s{x}[i] \approx \s{y}[i]$, for all $1\leq i \leq |\s{x}|$.  Given strings $\s{x}[1..n]$ and $\s{p}[1..m]$, \itbf{pattern matching} is the process of determining the positions $i \in 1..n-m+1$ 
in \s{x} such that \s{p} matches $\s{x}[i..i+m-1]$.

Let $\Sigma_p = \Sigma \cup *$, where $*$, called a \itbf{hole}~\cite{BS08}, matches every character in $\Sigma$. Then a string $\s{x}$ over $\Sigma_p$ that actually contains $*$ is called a \itbf{partial word}. Thus partial words are a special case of indeterminate strings in which every indeterminate letter is $\Sigma$.  
Early work on indeterminate strings dates back 30 or 40 years \cite{A87, FP74}, leading to many more recent papers studying their properties \cite{BS08, CIKMV16,DGGL19, HS03, SW08, SW09}.  Pattern-matching on indeterminate strings was studied early in the century \cite{HSW06,SW09} as well as more recently \cite{AABG20, CIKMV16, DGGL19, GILPP17, IR16}; the important special case of partial words has also attracted attention \cite{BS08,IMMS03,SW08}.  

In~\cite{HSW05,HSW06,HSW08}, Holub, Smyth and Wang describe a range of indeterminate algorithms with different restrictions imposed on text and pattern, the most general requiring $\O(mn)$ worst case execution time when both are indeterminate.  Versions of these algorithms were implemented and tested in \cite{HSW06} and \cite{SW09}, using a technique that flip flops between the Sunday variant of the Boyer-Moore algorithm~\cite{Sunday1990AVF} for regular strings and the Shift-Add algorithm in the indeterminate case \cite{BG92,D68,WM92}. Our algorithm \textsc{BM\_Indet}, described here, adapts the classical BM algorithm by using the prefix array for special strings to compute shifts --- a strategy different from the flip flop strategy adopted in~\cite{HSW05,HSW06,HSW08}.

In~\cite{IR16}, given indeterminate \s{x} and \s{p}, Iliopoulos and Radoszewski propose an algorithm based on the Fast Fourier Transform that executes in $2^{\sigma}$ steps, each requiring $\O(n\log m)$ time.  This is the best theoretical bound known so far, but the algorithm has not been implemented.  Pattern matching algorithms for \itbf{conservative} degenerate strings, where the number of indeterminate letters in text $\s{x}[1..n]$ and pattern $\s{p}[1..m]$ is bounded above by a constant $k$, have also been proposed~\cite{CIKMV16,DGGL19}.  In~\cite{CIKMV16}, Crochemore {\it et al.} describe an $\mathcal{O}(nk)$ algorithm which uses suffix trees and other auxiliary data structures to search for $\s{p}$ in $\s{x}$.  In~\cite{DGGL19}, Daykin {\it et al.} propose a pattern matching algorithm that first constructs the Burrows Wheeler Transform (BWT) of $\s{x}$ in $\mathcal{O}(n)$ time, then uses it to find all occurrences of any degenerate pattern $\s{p}$ in $\s{x}$ in time $O(mn)$. This drops to $\mathcal{O}(km^2 + q)$ when
the pattern $p$ is conservative, where $q$ is the number of occurrences of the pattern in $\s{x}$, and $\mathcal{O}(km^2)$ is the time required to compute them. 

Algorithms designed to process indeterminate strings sometimes encode the indeterminate string \s{x} into a 
match-preserving regular string \s{y}, then process \s{y} instead.  This encoding either maps the set-based alphabet $\Sigma'$ to a new alphanumeric alphabet, or maps the base alphabet $\Sigma$ to a numerical alphabet $\Sigma_{N}$.  The indeterminate letters are then processed based on some operation on the mapped or newly-constructed regular letters. For instance, the model proposed in~\cite{prochazka19} maps all the non-empty letters (both regular and indeterminate) over the DNA alphabet $\Sigma_{DNA}=\{A,C,G,T\}$ to $\Sigma_{IUPAC}=\{A,C,G,T,R,Y,S,W,K,M,B,D,H,V,N\}$ --- the 15 IUPAC letters. Then given $\s{x}$ over $\Sigma_{DNA}$, it constructs a corresponding $\s{y}$ over $\Sigma_{IUPAC}$. 

In~\cite{btt2013}, each letter of the DNA alphabet is mapped to a 4-bit integer power of $2$; that is, $\{A,C,G,T\}$ is mapped to $\Sigma_{N} = \{2^0, 2^1, 2^2, 2^3\}$. Then a non-empty indeterminate letter over $\Sigma_{DNA}$ is represented as $\Sigma_{\{s \in \mathcal{P}(\Sigma)\}}s$, again of size $15 = {1111}_2$. Also, instead of using the natural order on integers, \cite{btt2013} uses a Gray code~\cite{gray53} (also known as reflected binary Gray code) to order indeterminate letters over $\Sigma_{DNA}$. Since the Gray code ensures that two successive values differ by only one bit, such as 1000 and 1100, it becomes possible to minimize the number of separate intervals associated with each of the four symbols of $\Sigma_{DNA}$.

In this paper our contribution is two-fold.  We first propose a new encoding for indeterminate strings based on prime numbers which we use to describe two simple and space-efficient pattern matching algorithms, \textsc{KMP\_Indet} and \textsc{BM\_Indet}, derived from the classical KMP and BM algorithms, for indeterminate strings. Our algorithms avoid altogether the elaborate data structures used in previous algorithms \cite{CIKMV16,DGGL19, IR16}: Fast Fourier Transform, annotated suffix trees, Burrows Wheeler Transform.  Although both of the new algorithms require $O(nm^2)$-time in the worst case, we provide experimental evidence that \textsc{KMP\_Indet} nevertheless performs much faster in practice than the $O(n+km^2)$ algorithm described in \cite{DGGL19}, the only comparable algorithm whose working implementation is currently available\footnote{Though this implementation, while allowing indeterminate patterns, is restricted to texts required to be regular.}. 

The outline of the paper is as follows.  In Section~\ref{ssec:def} we give basic definitions and, for completeness, the classical \textsc{KMP} and \textsc{Boyer-Moore} algorithms. Section~\ref{sec:mapping} describes our prime number encoding for indeterminate strings. In Section~\ref{sec:patmat} we present two new algorithms, \textsc{KMP\_Indet} and \textsc{BM\_Indet}, for pattern matching on indeterminate strings. In Section~\ref{sec:exp} we compare the performance of four algorithms: 
\textsc{KMP\_Indet}, \textsc{BM\_Indet}, the brute force (\textsc{BF}) algorithm and the \textsc{DBWT} algorithm given in \cite{DGGL19}. Finally, Section~\ref{sec:conclusion} states some open problems and future directions for this style of pattern matching on indeterminate strings.

\section{Preliminaries}\label{ssec:def}
We give here a few essential definitions, and for completeness briefly describe the classical KMP and BM pattern matching algorithms on regular strings.

Given $\s{x}[1..n]$, for $1 \le i,j \le n$, $\s{u} = \s{x}[i..j]$ is called a \itbf{substring} of \s{x}, an \itbf{empty} substring \s{\varepsilon} if $j < i$.
If $i = 1$, \s{u} is a \itbf{prefix} of $\s{x}$, a \itbf{suffix} if 
$j = n$. A string $\s{x}$ (regular or indeterminate) is \itbf{periodic} if it can be written as $\s{x}=\s{u}^e\s{u}'$, \s{u'} a prefix of \s{x},
where either $e=1$ and $\s{u}' \neq \varepsilon$, or $e \geq 2$.

A string $\s{x}[1..n]$ has a \itbf{border} of length $t$ if $0 \le t < n$ and $\s{x}[1..t]$ matches $\s{x}[n-t+1..n]$.  Note that \s{x} always has an empty border. A \itbf{border array} $\beta_{\s{x}}=\beta_{\s{x}}[1..n]$ of $\s{x}$ is an integer array where for every $i \in [1..n]$, $\beta_{\s{x}}[i]$ is the length of the longest border of $\s{x}[1..i]$. A \itbf{prefix array} $\pi_{\s{x}}=\pi_{\s{x}}[1..n]$ of $\s{x}$ is an integer array where for every $i \in [1..n]$, $\pi_{\s{x}}[i]$ is the length of the longest substring starting at position $i$ that matches a prefix of $\s{x}$. See Figure~\ref{fig:betapiex} for an example. 

\begin{figure}
\begin{center}
\begin{tabular}{rccccccccccccc}
\hspace{7mm}& 1\hspace{0mm} & 2\hspace{0mm} & 3\hspace{0mm} & 4\hspace{0mm} & 5\hspace{0mm} & 6\hspace{0mm} & 7\hspace{0mm} & 8\hspace{0mm} & 9\hspace{0mm} & 10\hspace{0mm} & 11\hspace{0mm} & 12\hspace{0mm} & 13\\ [0.5ex] 
\s{x}\hspace{7mm} & a\hspace{0mm} & a\hspace{0mm} & b\hspace{0mm} & a\hspace{0mm} & a\hspace{0mm} & b\hspace{0mm}  & a\hspace{0mm} & a\hspace{0mm} & \{a, b\}\hspace{0mm} & b\hspace{0mm} & a \hspace{0mm} & a\hspace{0mm} & \{a, c\}\hspace{0mm}\\[0.5ex]
$\beta_{\s{x}}$\hspace{7mm} & 0\hspace{0mm} & 1\hspace{0mm} & 0\hspace{0mm} & 1\hspace{0mm} & 2\hspace{0mm} & 3\hspace{0mm} & 4\hspace{0mm} & 5\hspace{0mm} & \ \ 6\hspace{0mm}\hspace{0mm} & 3\hspace{0mm} & 4\hspace{0mm} & 5\hspace{0mm} & 2\hspace{0mm} \\[0.5ex]
$\pi_{\s{x}}$\hspace{7mm} & 13\hspace{0mm} & 1\hspace{0mm} & 0\hspace{0mm} & 6\hspace{0mm} & 1\hspace{0mm} & 0\hspace{0mm} & 3\hspace{0mm} & 5\hspace{0mm} & \ \ 1\hspace{0mm}\hspace{0mm} & 0\hspace{0mm} & 2\hspace{0mm} & 2\hspace{0mm} & \ 1\hspace{0mm} \\
\end{tabular}
\end{center}	
\caption{Border array $\beta_{\s{x}}$ and prefix array $\pi_{\s{x}}$ for string $\s{x}=aabaabaa\{a,b\}baa\{a, c\}$.}\label{fig:betapiex}
\end{figure}

In Lemmas~\ref{lem:ba} and~\ref{lem:prefixa}, we rephrase earlier results on running times for computing the border array and prefix array of a string of length $n$.

\begin{lemm}[\cite{AHU74,SW08}]\label{lem:ba}
The border and prefix arrays of a regular string of length $n$ can each be computed in $\mathcal{O}(n)$ time.
\end{lemm}
\begin{lemm}[\cite{S03,SW08}]\label{lem:prefixa}
The border and prefix arrays of an indeterminate string of length $n$, over a constant-sized alphabet, can each be computed in $\mathcal{O}(n^2)$ time in the worst case, $\mathcal{O}(n)$ time in the average case.
\end{lemm}

An improved upper bound for computing the prefix array is given in~\cite{IR16}:

\begin{lemm}[\cite{IR16}]\label{lem:prefixac}
The prefix array of an indeterminate string of length $n$ over a constant-sized alphabet can be computed in $\mathcal{O}(n \sqrt{n})$ time and $\mathcal{O}(n)$ space.
\end{lemm}

Given a (regular or indeterminate) string $\s{x}[1..n]$ and a (regular or indeterminate) pattern $\s{p}[1..m]$, $m\leq n$, the pattern matching problem requires computing all the index positions at which $\s{p}$ occurs in $\s{x}$.
Algorithms~\ref{fig:kmpreg} and~\ref{fig:bmreg} outline the Knuth Morris Pratt (\textsc{KMP}) \cite{KMP77} and Boyer Moore (\textsc{BM}) \cite{BM77} pattern matching algorithms for regular strings \s{x} and \s{p}.  

\textsc{KMP}, in case of a mismatch or after a full match, computes the shift of the pattern $\s{p} = \s{p}[1..m]$ along \s{x} by using the border array of \s{p}, which as we have seen is computable in $\mathcal{O}(m)$ time. Thus \textsc{KMP} runs in $\mathcal{O}(n + m)$ time using $\mathcal{O}(m)$ additional space.

\begin{algorithm}[!ht]
	\caption{KMP Algorithm}\label{fig:kmpreg}
	\begin{algorithmic}[1]
		\Function{KMP}{$\s{x},n,\s{p},m$}: Integer\text{ }List
            \State $i \la 0 ;\ j \la 0$
            \State $indexlist \la \emptyset$ \com{List of indices where $\s{p}$ occurs in $\s{x}$}
            \State $\beta_{\s{p}} \la $ Border array of pattern $\s{p}$
            \While {$i < n$}
                \If {$\s{p}[j+1] = \s{x}[i+1]$} 
                    \State $j \la j + 1;\ i \la i + 1$
                    \If {$j = m$}
                        \State $indexlist \la indexlist \cup \{i\m j \+ 1\}$
                        \State $j \la \beta_{\s{p}}[j]$
                    \EndIf
                \Else 
                    \If {$j=0$}
                        \State $i \la i \+ 1$ 
                    \Else
                        \State $j \la \beta_{\s{p}}[j]$
                    \EndIf
                \EndIf
            \EndWhile
        \State return $indexlist$
        \EndFunction
	\end{algorithmic}
\end{algorithm}

\begin{algorithm}[!ht]
	\caption{BM Algorithm}\label{fig:bmreg}
	\begin{algorithmic}[1]
		\Function{BM}{$\s{x},n,\s{p},m$}: Integer\text{ }List 
            \State $i \la 1 ; j \la m ; shift \la 1;$ $ \text{mismatched} \la \false$
             \State $indexlist \la \emptyset$ \Comment{List of indices where $\s{p}$ occurs in $\s{x}$}
             \While {$i < n-m+1$}
                \State $shift \la 1 ;\ mismatched \la \false$
                \For {$j \la m$ \bdownto $1$}
                     \If {$\s{p}[j] \neq \s{x}[i \+ j \m 1]$}
                        \State $skip\_bc \la$ \textsc{bad\_character\_rule\_shift}
                        \State $skip\_gs \la$ \textsc{good\_suffix\_rule\_shift}
                        \State $shift = MAX(shift, skip\_bs, skip\_gs)$
                        \State $mismatched \la \true$
                        \State break;
                    \EndIf
                \EndFor
                    \If {$!mismatched$}
                        \State $indexlist \la indexlist \cup \{i\}$
                        \State $skip\_gs \la$ \textsc{good\_suffix\_rule\_shift}
                        \State $shift = MAX(shift, skip\_gs)$
                    \EndIf
                \State $i = i \+ shift$
            \EndWhile
            
            \State return $indexlist$
        \EndFunction
	\end{algorithmic}
\end{algorithm}

On the other hand, the Boyer-Moore algorithm, in case of a full match or a mismatch of \s{p} in \s{x}, uses \itbf{bad character} and 
\itbf{good suffix} rules to compute the shift of \s{p} along \s{x}.  At each alignment, BM compares letters in the pattern right-to-left rather then KMP's left-to-right.
If during an alignment BM identifies a non-empty suffix $\s{s}$ of $\s{p}$ that matches a substring $\s{u}$ of $\s{x}$,
then by the \itbf{good suffix rule} the pattern is shifted  right till one of the following holds:
\begin{enumerate}
    \item The pattern moves past the occurrence of $\s{u}$ in the text (this happens when $\s{s}$ does not occur in $\s{p}[1..|\s{p}| \m |\s{s}|]$); or
    \item The pattern moves to the right until $\s{u}$ in the text aligns with the rightmost occurrence of itself in $\s{p}[1..|\s{p}|\m |\s{s}|]$; or
    \item The pattern moves to the right until a non-empty suffix of $\s{u}$ in the text aligns with the rightmost occurrence of itself in $\s{p}[1..|\s{p}|\m |\s{s}|]$.
\end{enumerate}

However, in case of a mismatch or after a full match during an alignment, the \itbf{bad character rule} is used: the mismatched character in the text (say $\alpha = \s{x}[i]$) is recorded and the pattern is shifted right until the rightmost occurrence of $\alpha$ in $\s{p}[1..i-1]$ aligns with $\s{x}[i]$, if it exists; otherwise, the pattern is shifted past $\s{x}[i]$. 

BM then uses the maximum shift returned by the good suffix and bad character rules to determine the final shift.  Using the bad character rule,  the shift can be computed in constant time, based on a table $D$, called a \itbf{bad character table}, pre-computed in $O(m\sigma)$ time, where $D[i,j]$ gives the rightmost occurrence of the character $i$ in the prefix of the pattern of length $j$. For the good suffix rule, a prefix table for the reverse of the pattern is constructed in $O(m)$ time to enable the shift to be computed in constant time.  Hence BM executes in at most 
$O(n + m\sigma)$ time and $O(m\sigma)$ space.
See \cite[Chs. 7--8]{S03} for a detailed discussion of these two algorithms and their variants.
The bad character rule is usually implemented using a one-dimensional table requiring $O(\sigma)$ space that is
 pre-computed in $O(m+\sigma)$ time,
 so that BM runs in $O(n + m+\sigma)$ time and $O(\sigma)$ space.

\section{Indeterminate String Encoding}\label{sec:mapping}

In this section we propose a new encoding to transform an indeterminate string $\s{x}$ into a regular string $\s{y}$ on an integer alphabet, such that for every $1 \leq i, j \leq n$, $\s{x}[i] \approx \s{x}[j] \Longleftrightarrow \s{y}[i] \approx \s{y}[j]$.
This encoding is effective on small alphabets 
($|\Sigma| =\sigma \le 9$), including in particular the important case of DNA sequences ($\Sigma_{DNA} = \{a,c,g,t\}$).
Thus cumbersome and time-consuming matches of indeterminate letters can be efficiently handled.

Let $P_{\sigma} = \{\mu_1, \mu_2, \ldots, \mu_{\sigma}\}$, where $\mu_j$ is the $j^{\mbox{th}}$ prime number ($\mu_1 = 2, \mu_2 = 3, ..., \mu_9 = 23$). We define the mapping $M: \Sigma \rightarrow P_{\sigma}$, where $\lambda_j \rightarrow \mu_j,\ j = 1,2,\ldots,\sigma$.  Then, given an indeterminate string $\s{x} = \s{x}[1..n]$ on $\Sigma$ (the \itbf{source string}), we can apply the mapping to compute a corresponding numerical regular string $\s{y} = \s{y}[1..n]$ (the \itbf{mapped string}) according to the following rule:

\begin{quote}
    (R) For every $\s{x}[i] = \{\lambda_{i1},\lambda_{i2},\ldots,\lambda_{ik}\}$, $1  \le ik \le \sigma$, $1 \le i \le n$, where $\lambda_{ih} \in \Sigma, 1 \le h \le k$,  set $$\s{y}[i] \la \prod^k_{h=1} M(\lambda_{ih}).$$
\end{quote}
When $k = \sigma$, \s{y} achieves the maximum value, which we denote by $M_{\sigma} = \prod^{\sigma}_{j=1} \mu_j$. More generally, since the mapping yields all possible products of the first $\sigma$ prime numbers, it imposes an order on indeterminate letters drawn from $\Sigma$:
$\s{x}[i_1] < \s{x}[i_2] \Leftrightarrow \s{y}[i_1] < \s{y}[i_2]$.

For example, consider a DNA source string $\s{x}= a\{a,c\}g\{a,t\}t\{c,g\}$, over $\Sigma_{DNA}$. Then $\sigma = 4$, and applying (R) for 
$1 \le k \le 4$ (based on the mapping $\mu: 2 \la a, 3 \la c, 5 \la g, 7 \la t$), we compute a mapped string $\s{y}=2/6/5/14/7/15$, so that
$$a < g < \{a,c\} < t < \{a,t\} < \{c,g\}.$$

\begin{lemm}
\label{lemm-0}
Rule (R) computes \s{y} in time $O(n)$ for a constant alphabet. 
\end{lemm}

\begin{lemm}
\label{lemm-1}
If \s{y} is computed from \s{x} by Rule (R), then for every $i_1,i_2 \in 1..n$, $\s{x}[i_1] \approx \s{x}[i_2]$ if and only if $\gcd(\s{y}[i_1],\s{y}[i_2]) > 1$.
\end{lemm}
\begin{proof}

\noindent($\Rightarrow$) By contradiction. Suppose $\s{x}[i_1] \approx \s{x}[i_2]$, $1 \leq i_1,i_2 \leq n$, but $\gcd(\s{y}[i_1],\s{y}[i_2]) = 1$; that is, $\s{y}[i_1]$ and $\s{y}[i_2]$ have no common divisor greater than one.  Since for every $i$, the letter $\s{y}[i]$ is a product of the prime numbers assigned to the characters in $\s{x}[i]$, we see that therefore $\s{x}[i_1]$ and $\s{x}[i_2]$ can have no character in common; that is, $\s{x}[i_1] \not \approx \s{x}[i_2]$, a contradiction. 

\noindent($\Leftarrow$) By the reverse argument.  \qed
\end{proof}

Two strings \s{x_1} and \s{x_2} of equal length $n$ are said to be \itbf{isomorphic} if and only if
for every $i,j \in$ $\{1,\dots,n\}$,
\begin{equation}
\label{rule2}
\s{x_1}[i] \approx \s{x_1}[j] \Longleftrightarrow \s{x_2}[i] \approx \s{x_2}[j].
\end{equation}

Thus we have:
\begin{obs}
If $\s{x}$ is an indeterminate string on $\Sigma$ and $\s{y}$ is the numerical string constructed by applying Rule (R) to $\s{x}$, then $\s{x}$ and $\s{y}$ are isomorphic. 
\end{obs}
\begin{obs}
By virtue of Lemma~\ref{lemm-1} and (\ref{rule2}), \s{y} can overwrite the space required for \s{x} (and {\it vice versa}) with no loss of information.
\end{obs}

\begin{obs}
\label{obs-gcd}
By~\cite[pp. 316--364]{KNUTH69II}, suppose $\ell_1$ and $\ell_2 < \ell_1$ are integers, where $\ell_2$ can be represented by $D = \log_{10} \ell_2$ decimal digits.
Then $\gcd(\ell_1,\ell_2)$ can be computed in time 
$\O(D)$.
\end{obs}
\begin{obs}
For $\sigma = 9$ corresponding to the first nine prime numbers  $$2,3,5,7,11,13,17,19,23,$$ the product $M_{\sigma} = 223,092,870$, a number representable in $D = 9$ decimal digits, also in $B = 32$ bits, a single computer word.  Thus by Observation~\ref{obs-gcd}, the time required to match any two indeterminate letters is proportional to $D$.  When $\sigma = 4$, corresponding to $\Sigma_{DNA}$, $2 \times 3 \times 5 \times 7 = 210 < 256$, and so $D = 4$, and the matching time is correspondingly reduced.
\end{obs}

To summarize: for $\sigma \le 9$, computing a match between $\s{x}[i_1]$ and $\s{x}[i_2]$ on $\Sigma$ (that is, between $\s{y}[i_1]$ and $\s{y}[i_2]$) requires time bounded above by a (small) constant.

\section{Pattern Matching Algorithms on Indeterminate Strings}\label{sec:patmat}

Here we describe two simple and space-efficient algorithms, \textsc{KMP\_Indet} and \textsc{BM\_Indet}, that, in order to find all occurrences of an indeterminate pattern $\s{p} = \s{p}[1..m]$ in an indeterminate string $\s{x} = \s{x}[1..n]$, compute all occurrences of a corresponding mapped pattern $\s{q} = \s{q}[1..m]$ in a mapped string $\s{y} = \s{y}[1..n]$.

\subsection{KMP Algorithm on Indeterminate Strings}\label{ssec:kmpindet}
\textsc{KMP\_Indet} (Algorithm~\ref{fig-KMPI}) searches for pattern $\s{q}=\s{q}[1..m]$ in text $\s{y}=\s{y}[1..n]$, outputting the indices at which $\s{q}$ occurs in $\s{y}$ (thus, at which \s{p} occurs in \s{x}). Our algorithm therefore implements the KMP algorithm \cite{KMP77} on indeterminate strings that have been transformed using Rule (R). Although this transformation is not necessary for the algorithm to work, we use it to improve space and time efficiency.  The algorithm also works with other indeterminate string encoding/transformations described in Section~\ref{sec:intro}. While scanning $\s{y}$ from left to right and performing letter comparisons, \textsc{KMP\_Indet} checks whether the prefix of $\s{q}$ and the substring of $\s{y}$ currently being matched are both regular. If so, then it uses the border array $\beta_{\s{q}_{\ell}}$ of the length-$\ell$ longest regular prefix $\s{q}_{\ell}$ of $\s{q}$ to compute the shift; if not, 
in case of a mismatch between $\s{q}[j+1]$ and $\s{y}[i+1]$, it constructs a new string
$\s{q'}=\s{q}[1..j-1]\s{y}[i-j+2..i]$, which is a concatenation of the longest proper prefix of the matched prefix of $\s{q}$ with the longest proper suffix of the matched substring of $\s{y}$. Then it constructs the prefix array $\pi_{\s{q'}}$ of $\s{q'}$ to compute the shift. The \textsc{Compute\_Shift} function given in Algorithm~\ref{fig-comshift} implements this computation.
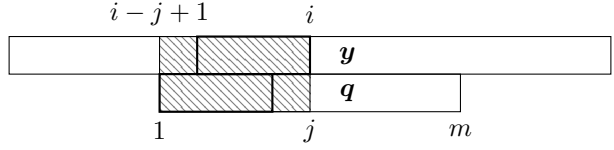
\begin{figure}
\begin{center}
\begin{tikzpicture}
\draw (-4,-0.25) rectangle (4, 0.25);
\draw[pattern=north west lines, pattern color=gray] (0,-0.25) rectangle (-2,0.25); 
\draw[pattern=north west lines, pattern color=gray] (0,-0.25) rectangle (-2,-0.75); 
\draw [thick] (0,-0.25) rectangle (-1.5,0.25); 
\draw [thick](-0.5,-0.25) rectangle (-2,-0.75); 
\draw (-2,-0.25) rectangle (2, -0.75);
\draw node at (0.5,0.0) {$\s{y}$};
\draw node at (0.5,-0.5) {$\s{q}$};
\draw node at (-2,.55) {$i-j+1$};
\draw node at (0,.55) {$i$};
\draw node at (-2,-1) {$1$};
\draw node at (0,-1) {$j$};
\draw node at (2,-1) {$m$};
\end{tikzpicture}
\end{center}
\caption{Illustration of the \textsc{KMP\_Indet} algorithm when a prefix of the pattern $\s{q}[1..j]$ matches the text substring $\s{y}[i-j+1..i]$. If either $\s{q}[1..j-1]$ or $\s{y}[i-j+2..i]$ contains an indeterminate letter, then we construct $\s{q'} = \s{q}[1..j-1]\s{y}[i-j+2..i]$ to compute the shift by constructing the prefix table for $\s{q'}$.}\label{fig:kmp-indet}
\end{figure}

In order to determine whether or not indeterminate letters are included in any segment $\s{q'} = \s{q}[1..j\m 1]\s{y}[i\m j \+ 2..i]$, two variables are employed: $indet_y$ and the length  $\ell$ of the longest regular prefix $\s{q_{\ell}}$ of \s{q}.
$indet_y$ is a Boolean variable that is $\true$ if and only if the current segment $\s{y}[i\m j \+ 2..i]$ contains an indeterminate letter; $\ell$ is pre-computed in $\O(m)$ time as a byproduct of the one-time calculation of $\s{q_{\ell}}$. 

\begin{algorithm}[!ht]
	\caption{KMP\_Indet Algorithm}\label{fig-KMPI}
	\begin{algorithmic}[1]
		\Function{\textsc{KMP\_Indet}}{$\s{y},n,\s{q},m$}: Integer\text{ }List 
            \State $i \la 0 ;\ j \la 0;\ indet_y \la \false$$ ;\ right\_pos=0$
            \State $indexlist \la \emptyset$ \Comment{List of indices where $\s{q}$ occurs in $\s{y}$}
            \State $\s{q}_{\ell} \la$ longest regular prefix of $\s{q}$ of length $\ell$
            \State $\beta_{\s{q}} \la Compute\_\beta(\s{q}_{\ell})$ \com{Border Array of $\s{q}_{\ell}$}
            \While {$i < n$}
                \If {$\s{q}[j+1] \approx \s{y}[i+1]$}
                    \If {INDET(\s{y}[i + 1])} $indet_y \la \true$$ ;\ right\_pos=i\+1$ \EndIf
                    \State $j \la j + 1;\ i \la i + 1$
                    \If {$j = m$} 
                        \State $indexlist \la indexlist \cup \{i\m j \+ 1\}$
                        \State $j \la \textsc{Compute\_Shift}(indet_y, \s{y}, \s{q}, i, j, \beta_{\s{q}},\ell)$
                        \State $set\_indet_y(i,j,right\_pos)$
                    \EndIf
                \Else 
                    \If {$j=0$} 
                        \State $i \la i \+ 1$ 
                    \Else
                        \State $j \la \textsc{Compute\_Shift}(indet_y, \s{y}, \s{q}, i, j, \beta_{\s{q}},\ell)$
                    \EndIf
                    \State $set\_indet_y(i,j,right\_pos)$
                \EndIf
            \EndWhile
             \State return $indexlist$
        \EndFunction
	\end{algorithmic}
\end{algorithm}

If $\s{y}$ and $\s{q}$ are both regular, \textsc{KMP\_Indet} reduces to the KMP algorithm \cite{KMP77}.  Otherwise, it checks whether any indeterminate letter exists in the matched prefix of $\s{q} = \s{q}[1..j\m 1]$, or in the matched substring of $\s{y}=\s{y}[i \m j \+ 2..i]$. If so, then the shift in $\s{q}$ is equal to the maximum length of the prefix of $\s{q}[1..j\m 1]$ that matches with a suffix of $\s{y}[i \m j \+ 2..i]$. To compute this length, the algorithm first builds a new string $\s{q'}=\s{q}[1..j\m 1]\s{y}[i\m j \+ 2..i]$ and, based on an insight given in \cite{SW08}, computes its {\it prefix} array $\pi_{\s{q'}}$ rather than its border array. To compute the shift only the last $j-1$ entries of $\pi_{\s{q'}}$ are examined; that is, entries $k=j$ to $2(j-1)$. Note that we need to consider only those entries $r$ in $\pi_{\s{q'}}[j..2(j-1)]$ where a prefix of $\s{q'}$ matches the suffix at $r$ ($\s{q'}[r..2(j-1)]$); that is, the entries where $\pi_{\s{q'}}[r] = 2j \m r \m 1$. The shift is simply the maximum over such entries in $\pi_{\s{q'}}$. Recall that computing the border array for an indeterminate string is not useful as the matching relation $\approx$ is not transitive \cite{HS03}. Finally, the $set\_indet_y(i,j,right\_pos)$ function checks if any indeterminate letter exists in the substring $\s{y}[i-j+1..j]$. If it does it sets $indet_y$ to $\true$; otherwise it sets it to $\false$.

\begin{algorithm}[h]
	\caption{Compute\_Shift Algorithm}\label{fig-comshift}
	\begin{algorithmic}[1]
		\Function{\textsc{Compute\_Shift}}{$(indet_y, \s{y}, \s{q}, i, j, \beta_{\s{q}},\ell$}: Integer
  \Comment{$\ell$ is length of longest regular prefix of \s{q}.}
  \If{$indet_y$ \bor $j > \ell$}
    \State  $\s{q'} \la \s{q}[1..j\m 1]\s{y}[i \m j \+ 2..i]$
     \If{$\s{q'} = \varepsilon$}
       \State return $0$
     \EndIf
     \State $\pi_{\s{q'}} \la Compute\_\pi(\s{q'})$ \Comment{Prefix Array of pattern $\s{q'}$}
     \State $max \la 0$
     \For{$r \la j$ \bto $2(j-1)$}
       \If{$max < \pi_{\s{q'}}[r]$ \band $\pi_{\s{q'}}[r] = 2j \m r \m 1$}
         \State $max \la \pi_{\s{q'}}[r]$
       \EndIf
     \EndFor
     \State $j \la max$
  \Else \Comment{prefix of $\s{q}$ \& substring of $\s{y}$ are regular}
     \State $j \la \beta_{\s{q}}[j]$
  \EndIf
  \State return $j$
	        \EndFunction
	\end{algorithmic}
\end{algorithm}

Figure~\ref{fig:kmpindetex} represents the processing of the text $\s{x}=aabaabaa\{a,b\}baa\{a,c\}$ and pattern $\s{p}=aabaa$ corresponding to the processing of \s{y} and \s{q} by \textsc{KMP\_Indet}. \textsc{KMP\_Indet} first computes $\beta_{\s{p}}=(0, 1, 0, 1, 2)$ and $\ell = 5$. Initially the pattern is aligned with $\s{x}$ at position $1$. Since it matches with the text ($j=5$), and $indet_x=\false$ and $5 \leq (\ell = 5)$, we compute the shift from $\beta_{\s{p}}[5]=2$. Thus the pattern is aligned with $\s{x}$ at position $i=4$. Analogously, the pattern is next aligned with $\s{x}$ at position $i=7$. Since a mismatch occurs at $i+1=10, j+1 = 4$, and because $indet_x = \true$, we construct $\s{p}' = \s{p}[1..2]\s{x}[8..9]=aaa\{a,b\}$ and compute $\pi_{\s{p'}}=(4, 3 , 2, 1)$. Thus the shift is equal to $2$ and so the pattern is aligned with $\s{x}$ at position $8$. Since it matches (and because it is the last match), \textsc{KMP\_Indet} returns the list $\{1, 4, 8\}$.

\textsc{KMP\_Indet} contains a function INDET that determines whether or not the current letter $\s{y}[i\+ 1]$ is indeterminate. To enable this query to be answered in constant time, for $\sigma = 9$, we suppose that an array $P = P[1..23]$ has been created with $P[t] = -1$ if $t$ is a prime; otherwise $P[t]=1$.  Then $\s{y}[i\+ 1]$ is indeterminate if and only if it exceeds 23 or $P[\s{y}[i\+ 1]] = 1$.

\begin{figure}
\begin{center}
\begin{tabular}{rccccccccccccc}
\hspace{7mm}& 1\hspace{0mm} & 2\hspace{0mm} & 3\hspace{0mm} & 4\hspace{0mm} & 5\hspace{0mm} & 6\hspace{0mm} & 7\hspace{0mm} & 8\hspace{0mm} & 9\hspace{0mm} & 10\hspace{0mm} & 11\hspace{0mm} & 12\hspace{0mm} & 13\\ [0.5ex] 

\s{x}\hspace{7mm} & a\hspace{0mm} & a\hspace{0mm} & b\hspace{0mm} & a\hspace{0mm} & a\hspace{0mm} & b\hspace{0mm}  & a\hspace{0mm} & a\hspace{0mm} & \{a, b\}\hspace{0mm} & b\hspace{0mm} & a \hspace{0mm} & a\hspace{0mm} & \{a, c\}\hspace{0mm}\\[0.5ex]

& a\hspace{0mm} & a\hspace{0mm} & b\hspace{0mm} & a\hspace{0mm} & a\hspace{0mm} &\hspace{0mm}&\hspace{0mm}&\hspace{0mm}&\hspace{0mm}&\hspace{0mm}&\hspace{0mm}&\hspace{0mm}\\[0.5ex]

& \hspace{0mm} & \hspace{0mm} & \hspace{0mm} & a\hspace{0mm} & a\hspace{0mm} &b\hspace{0mm}&a\hspace{0mm}&a\hspace{0mm}&\hspace{0mm}&\hspace{0mm}&\hspace{0mm}&\hspace{0mm}\\[0.5ex]

& \hspace{0mm} & \hspace{0mm} & \hspace{0mm} & \hspace{0mm} & \hspace{0mm} &\hspace{0mm}&a\hspace{0mm}&a\hspace{0mm}&b\hspace{0mm}&x\hspace{0mm}&\hspace{0mm}&\hspace{0mm}\\[0.5ex]

& \hspace{0mm} & \hspace{0mm} & \hspace{0mm} & \hspace{0mm} & \hspace{0mm} &\hspace{0mm}&\hspace{0mm}&a\hspace{0mm}&a\hspace{0mm}&b\hspace{0mm}&a\hspace{0mm}&a\hspace{0mm}\\[0.5ex]

\end{tabular}
\end{center}	
\caption{The figure simulates the execution of \textsc{KMP\_Indet} on the text $\s{x}=aabaabaa\{a,b\}baa\{a,c\}$ and pattern $\s{p}=aabaa$. After execution, \textsc{KMP\_Indet} returns the list of positions $\{1, 4, 8\}$ at which $\s{p}$ occurs in $\s{x}$. `x' in the third alignment identifies a mismatch.} \label{fig:kmpindetex}
\end{figure}

Now we discuss the running time of algorithm \textsc{KMP\_Indet}. Clearly, for regular pattern and regular text, the time is linear. Otherwise, when a matched prefix of $\s{q}$ or a matched substring of $\s{y}$ contains an indeterminate letter, then the algorithm constructs the prefix array of a new string $\s{q}'$. In the worst case we might need to construct the prefix array of $\s{q'}$ for each iteration of the \bwhile loop. By Lemma~\ref{lem:prefixac} and because $\s{q'}$ can be of length at most $2(m\m 1)$, in the worst case the total time required for the execution of \textsc{KMP\_Indet} is $\mathcal{O}(nm\sqrt m)$. 
Theorem~\ref{thm:runKMPIc} states these conclusions:

\begin{thrm}\label{thm:runKMPIc}
Given text $\s{y}=\s{y}[1..n]$ and pattern $\s{q}=\s{q}[1..m]$ on a constant alphabet of size $\sigma$,  \textsc{KMP\_Indet} executes in $\mathcal{O}(n)$ time when $\s{y}$ and $\s{q}$ are both regular; otherwise, when at least one of them is indeterminate, the worst-case upper bound is 
$\mathcal{O}(nm\sqrt{m})$.  The algorithm's additional space requirement is $\O(m)$, for the pattern \s{q'} and corresponding arrays $\beta_{\s{q'}}$ and $\pi_{\s{q'}}$.
\end{thrm}

\begin{remrk}
Note that, apart from the $\O(n)$ time in-place mapping of \s{x} into \s{y} and \s{p} into \s{q}, there is no preprocessing and the only auxiliary time/space requirement relates to the use of the prefix array or border array to compute shifts of (usually short) substrings of \s{y}.  As a result,  \textsc{KMP\_Indet} processing is direct and immediate, requiring little additional storage.
\end{remrk}
\begin{remrk}
The worst case time requirement is predicated on a requirement for $\O(n)$ (short) shifts of \s{q} along \s{y}, each requiring by Lemma~\ref{lem:prefixac} a worst-case $\mathcal{O}(m \sqrt{m})$ prefix array calculation.
For example, this circumstance could occur with 
$\s{p} =  \{a,b\}c^{m-1}$ and $\s{x} = a^n$ or with
$\s{p} = ab$ and $\s{x} = \{a,c\}^n$.
\end{remrk}

\subsection{BM Algorithm on Indeterminate Strings}
\label{ssec:bmindet}

In this section we describe the \textsc{BM\_Indet} algorithm (see Algorithm~\ref{fig:bmi}). Like \textsc{KMP\_Indet} it searches for pattern $\s{q}=\s{q}[1..m]$ in text $\s{y}=\s{y}[1..n]$ and outputs the indices at which $\s{q}$ occurs in $\s{y}$. However, similar to \textsc{BM}, while scanning $\s{y}$ from left to right, at each alignment it performs letter comparisons from right to left in the pattern. If a mismatch occurs or if matching extends beyond the end of the pattern, the next shift is computed based on the maximum of those computed: first, from the bad character rule extended to indeterminate characters and, second, from the modified good suffix rule.

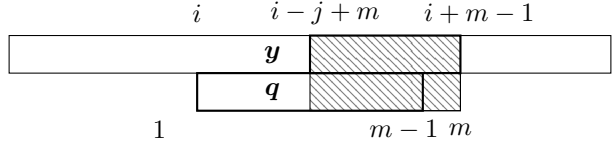
\begin{figure}
\begin{center}
\begin{tikzpicture}
\draw (-4,-0.25) rectangle (4, 0.25);
\draw[pattern=north west lines, pattern color=gray] (0,-0.25) rectangle (2,0.25); 
\draw[pattern=north west lines, pattern color=gray] (0,-0.25) rectangle (2,-0.75); 
\draw [thick] (0.0,-0.25) rectangle (2,0.25); 
\draw [thick](1.5,-0.25) rectangle (-1.5,-0.75); 
\draw (-1.5,-0.25) rectangle (2, -0.75);
\draw node at (-0.5,0.0) {$\s{y}$};
\draw node at (-0.5,-0.5) {$\s{q}$};
\draw node at (2.25,.55) {$i+m-1$};
\draw node at (0.2,.55) {$i-j+m$};
\draw node at (-1.5,.55) {$i$};
\draw node at (-2,-1) {$1$};
\draw node at (1.25,-1) {$m-1$};
\draw node at (2,-1) {$m$};
\end{tikzpicture}
\end{center}
\caption{Illustration of the \textsc{BM\_indet} algorithm when a suffix of the pattern of length $j \m 1$ $\s{q}[j-1..m]$ matches the text substring $\s{y}[i+j-2..i+m-1]= \s{t}'$. If either $\s{q}[1..m-1]$ or $\s{t}'$ contains an indeterminate letter, then we construct $\s{q'} = \s{t'}^R\s{q}[1..m-1]^R$ to compute the shift. }\label{fig:bm-indet}
\end{figure}

\begin{algorithm}[!ht]
	\caption{BM\_Indet Algorithm}\label{fig:bmi}
	\begin{algorithmic}[1]
		\Function{\textsc{BM\_Indet}}{$\s{y},n,\s{q},m$}: Integer\text{ }List 
            \State $i \la 1;\ shift \la 1;\ mismatched \la \false$$;\ indet_{\s{y}} \la \false$
            \State $\ell \la \texttt{length of the longest regular suffix of } \s{q}$ 
            \State $indexlist \la \emptyset$ \Comment{List of indices where $\s{q}$ occurs in $\s{y}$}
            \While {$i < n-m+1$}
                \State $shift \la 1 ;\ mismatched \la \false$
                \For {$j \la m$ \textbf{downto} $1$}
                   \If {INDET($\s{y}[i \+ j \m1]$)}
                     $indet_y \la \true$
                    \EndIf
                    \If {$\s{q}[j] \neq \s{y}[i \+ j \m 1]$}
                        \State $skip\_bc \la$ \textsc{bad\_character\_rule\_shift}
                        \State $j \la j-1$ 
                        \If {$indet_y$ OR $m \m j \+ 1 > \ell$} 
                            \State $skip\_gs \la$ \textsc{indet\_gsr\_shift$(\s{y}, \s{q}, i, m \m j \+ 1)$}  
                        \Else
                            \State $skip\_gs \la$ \textsc{good\_suffix\_rule\_shift}
                        \EndIf
                        \State $shift = MAX(shift, skip\_bs, skip\_gs)$
                        \State $mismatched \la \true$
                        \State \bbreak
                    \EndIf
                \EndFor
                \If {$!mismatched$}
                    \State $indexlist \la indexlist \cup \{i\}$
                    \If {$indet_y$ OR $m \m j \+ 1 > \ell$}
                        \State $skip\_gs \la$ \textsc{indet\_gsr\_shift$(\s{y}, \s{q}, i, m)$}
                    \Else
                        \State $skip\_gs \la$ \textsc{good\_suffix\_rule\_shift}
                    \EndIf
                    \State $shift = MAX(shift, skip\_gs)$
                \EndIf
                \State $i = i \+ shift$
            \EndWhile
        \State return $indexlist$
        \EndFunction
	\end{algorithmic}
\end{algorithm}

 While scanning $\s{y}$ from left to right and performing letter comparisons from right to left, \textsc{BM\_Indet} checks whether $\s{q}[1..m-1]$ and the matched substring of length $j$ of $\s{y}$ are both regular. If so, it computes the shift based on the classical Boyer-Moore approach; otherwise, to compute the shift applying the good suffix rule, as shown in Figure~\ref{fig:bm-indet}, it constructs a new string $\s{q'}$ from the reverse of the matched substring in $\s{y}$, and the reverse of the prefix $\s{q}[1..m-1]$ of $\s{q}$. Hence, if $\s{t}'=\s{y}[i+j-1..i+m-1]$ is the matched substring in $\s{y}$
 then $\s{q}'= \s{t}'[1..j]^R\s{q}[1..m-1]^R$. Then we compute the prefix array $\pi_{\s{q'}}$ of $\s{q'}$, to compute the rightmost occurrence of $\s{t}'$ (or its longest suffix) in $\s{q}[1..m-1]$, and compute the shift accordingly. 

\begin{algorithm}[!ht]
	\caption{INDET\_GSR\_SHIFT}\label{fig-comshiftgsr}
	\begin{algorithmic}[1]
		\Function{\textsc{Indet\_gsr\_shift}}{$\s{y},\s{q},i, matchedlen$}: Integer\text{ } 
        \State $t\_len \la matchedlen$
        \State $\s{q'} \la \s{y}[i-t\_len+m..i+m-1]^R\s{q}[1..m-1]^R$
        \State $\pi_{\s{q'}} \la Compute\_\pi(\s{q'})$ \com{Prefix Array of pattern $\s{q'}$}
        \State $rindex \la t\_len + 1$ 
        \For {$k \la t\_len + 1$ \bto $|\s{q'}|$}
             \If {$\pi_{\s{q'}}[k] = t\_len$}
                \State $rindex \la k; \bbreak$ 
            \EndIf
                \If {$\pi_{\s{q'}}[k] > \pi_{\s{q'}}[rindex]$ \band $\pi_{\s{q'}}[k] < t\_len$}
                    \State $rindex \la k$
                \EndIf
        \EndFor
        \State $gs\_shift \la m \m (|\s{q'}|-rindex + 1)$
        \State return $gs\_shift$
        \EndFunction
	\end{algorithmic}
\end{algorithm}

 The \textsc{indet\_gsr\_shift} function given in Algorithm~\ref{fig-comshiftgsr} computes the shift. For this calculation, only the last $m \m 1$ entries of $\pi_{\s{q'}}$ are examined, as we are interested only in the occurrence of the matched substring (or its longest suffix) in the prefix of the pattern $\s{q}[1..m\m 1]$. Then the shift is the leftmost occurrence of the maximum which is $\leq |\s{t}'|$ over all the last $m-1$ entries in $\pi_{\s{q'}}$.

In order to determine whether $\s{q}'$ contains indeterminate letters, we employ exactly the same two variables, $indet_y$ and $\ell$, introduced in Section \ref{ssec:kmpindet} for  \textsc{KMP\_Indet}. However, in this case, $\ell$ is the length of the longest regular suffix $\s{q_{\ell}}$ of \s{q}.

\begin{figure}
\begin{center}
\begin{tabular}{rccccccccccccc}
\hspace{7mm}& 1\hspace{0mm} & 2\hspace{0mm} & 3\hspace{0mm} & 4\hspace{0mm} & 5\hspace{0mm} & 6\hspace{0mm} & 7\hspace{0mm} & 8\hspace{0mm} & 9\hspace{0mm} & 10\hspace{0mm} & 11\hspace{0mm} & 12\hspace{0mm} & 13\\ [0.5ex] 

\s{x}\hspace{7mm} & a\hspace{0mm} & a\hspace{0mm} & b\hspace{0mm} & a\hspace{0mm} & a\hspace{0mm} & b\hspace{0mm}  & a\hspace{0mm} & a\hspace{0mm} & \{a, b\}\hspace{0mm} & b\hspace{0mm} & a \hspace{0mm} & a\hspace{0mm} & \{a, c\}\hspace{0mm}\\[0.5ex]

& a\hspace{0mm} & a\hspace{0mm} & b\hspace{0mm} & a\hspace{0mm} & a\hspace{0mm} &\hspace{0mm}&\hspace{0mm}&\hspace{0mm}&\hspace{0mm}&\hspace{0mm}&\hspace{0mm}&\hspace{0mm}\\[0.5ex]

& \hspace{0mm} & \hspace{0mm} & \hspace{0mm} & a\hspace{0mm} & a\hspace{0mm} &b\hspace{0mm}&a\hspace{0mm}&a\hspace{0mm}&\hspace{0mm}&\hspace{0mm}&\hspace{0mm}&\hspace{0mm}\\[0.5ex]

& \hspace{0mm} & \hspace{0mm} & \hspace{0mm} & \hspace{0mm} & \hspace{0mm} &\hspace{0mm}&-\hspace{0mm}&-\hspace{0mm}&-\hspace{0mm}&x\hspace{0mm}&a\hspace{0mm}&\hspace{0mm}\\[0.5ex]

& \hspace{0mm} & \hspace{0mm} & \hspace{0mm} & \hspace{0mm} & \hspace{0mm} &\hspace{0mm}&\hspace{0mm}&a\hspace{0mm}&a\hspace{0mm}&b\hspace{0mm}&a\hspace{0mm}&a\hspace{0mm}\\[0.5ex]

\end{tabular}
\end{center}	
\caption{The figure simulates the execution of \textsc{BM\_Indet} on the text $\s{x}=aabaabaa\{a,b\}baa\{a,c\}$ and pattern $\s{p}=aabaa$. After execution, \textsc{BM\_Indet} returns the list of positions $\{1, 4, 8\}$ at which $\s{p}$ occurs in $\s{x}$. `x' in the third alignment identifies a mismatch.} \label{fig:bmindetex}
\end{figure}

Figure~\ref{fig:bmindetex} describes the processing of text 
$\s{x} = aabaabaa\{a,b\}baa\{a,c\}$
and pattern $\s{p}=aabaa$ corresponding to the processing of \s{y} and \s{q} by \textsc{BM\_Indet}. \textsc{BM\_Indet} first computes $\ell = 5$ and table $D$ to compute the shift using the bad character rule.
Initially the pattern is aligned with $\s{y}$ at position $1$. Since this alignment results in a match, $j=0$ and $indet_{\s{y}} = \false$, yielding $shift=3$ from the good suffix rule of the classical BM algorithm.
Then the pattern is aligned with $\s{y}$ at position $i=4$. Similarly, the pattern is next aligned with $\s{y}$ at position $i=7$. 
Since a mismatch occurs at $i=7$, we set $j = 4$, and because $indet_{\s{y}}=\false$, by the classical BM algorithm, we set $shift=1$. The pattern is then aligned with $\s{y}$ at position $i=8$. 
Since this alignment results in a match, defined  by $j=0$ and $indet_{\s{y}}=\true$, we compute the shift by constructing $\s{q}'= aab\{a,b\}abaa$ for which $\pi_{\s{q'}}=(8,1,0,5,1,0,2,1)$. Thus the shift equals $5-2=3$. Therefore the pattern is aligned with $\s{y}$ at $11$. Since the suffix starting at $11$ is smaller than the length of the pattern, \textsc{BM\_Indet} returns the list $\{1, 4, 8\}$.

By a run time analysis analogous to that given for \textsc{KMP\_Indet}, we have Theorem~\ref{thm:runBMI} that gives the time complexity of \textsc{BM\_Indet}.

\begin{thrm}\label{thm:runBMI}
Given text $\s{y}=\s{y}[1..n]$ and pattern $\s{q}=\s{q}[1..m]$ on a constant alphabet of size $\sigma$,  \textsc{BM\_Indet} executes in $\mathcal{O}(n)$ time when $\s{y}$ and $\s{q}$ are both regular; otherwise, when both are indeterminate, the worst-case upper bound is 
$\mathcal{O}(nm\sqrt{m})$. The algorithm's additional space requirement is $\O(m)$, for the pattern \s{q'}, its prefix array $\pi_{\s{q'}}$, and the bad character table.
\end{thrm}

 \begin{figure}[ht!]
    \centering
	\includegraphics[scale=.4]{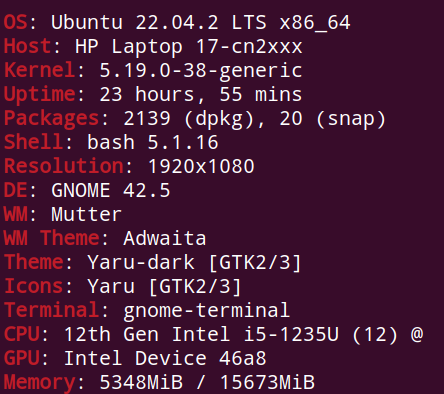}
	\caption{Specification of the server used to conduct experiments.} 
	\label{fig:serverspec}
\end{figure}

\section{Experimental Results}\label{sec:exp}

In this section we first present results of our experiments comparing the performance of four algorithms --- \textsc{KMP\_Indet}, \textsc{BM\_Indet}, \textsc{DBWT}~\cite{DGGL19} and brute force (\textsc{BF}) --- on randomly-generated indeterminate strings. We discover circumstances where each of these algorithms performs better than the others. The \textsc{DBWT} algorithm's implementation accepts strings only over four letter alphabet; that is, $\sigma = 4$. Moreover, it also requires that the text be regular. Given this, to enable us to perform additional testing of our algorithms, such as, on strings over varied alphabet sizes, on strings with varying indeterminate letter count in both text and pattern, and vary text and pattern sizes, we exclude \textsc{DBWT} in our further comparisons.
         
To improve the performance of \textsc{KMP\_Indet} and \textsc{BM\_Indet} several optimizations were included in the implementation. Furthermore, the \textsc{BM\_Indet} algorithm was implemented by constructing the prefix array from right to left using the approach given in~\cite{SW08}.

All our algorithms have been implemented in C++\footnote{The implementation code can be found at: \url{https://github.com/dehhganii/Practical_KMP_BM_Indet/tree/main}.}. The experiments were performed on a machine with the specifications given in Figure~\ref{fig:serverspec}. The test data was randomly generated using Python based on the numerical encoding described in Section~\ref{sec:mapping}, covering various alphabet sizes ($\sigma = 2, 3, 4, 9$), as well as varying string lengths, numbers of indeterminate letters, and positions of occurrence in the strings. 

In the graphs, the X-axis represents increasing string length and the Y-axis represents the running time (in seconds) of the algorithms. For text length $n=1000\times i$, we refer to \itbf{short} texts when $1 \leq i \leq 10$ and \itbf{long} texts when $100 \leq i \leq 1000$. The number of indeterminate letters in the text and pattern are denoted by $k_1$ and $k_2$, respectively.

\subsection{Comparison of \textsc{KMP\_Indet}, \textsc{BM\_Indet}, \textsc{DBWT}, and \textsc{BF} Algorithms}\label{ssec:comp1}

In this section we perform several experiments to discover scenarios where each algorithm performs best. Note that, since \textsc{DBWT} algorithm's implementation only accepts regular texts and strings over a four letter alphabet, we set $\sigma = 4$ and $k_1=0$.

 \begin{figure}[!ht]
         \centering 
         \begin{minipage}{.45\textwidth}
                \centering
               \includegraphics[scale=.175]{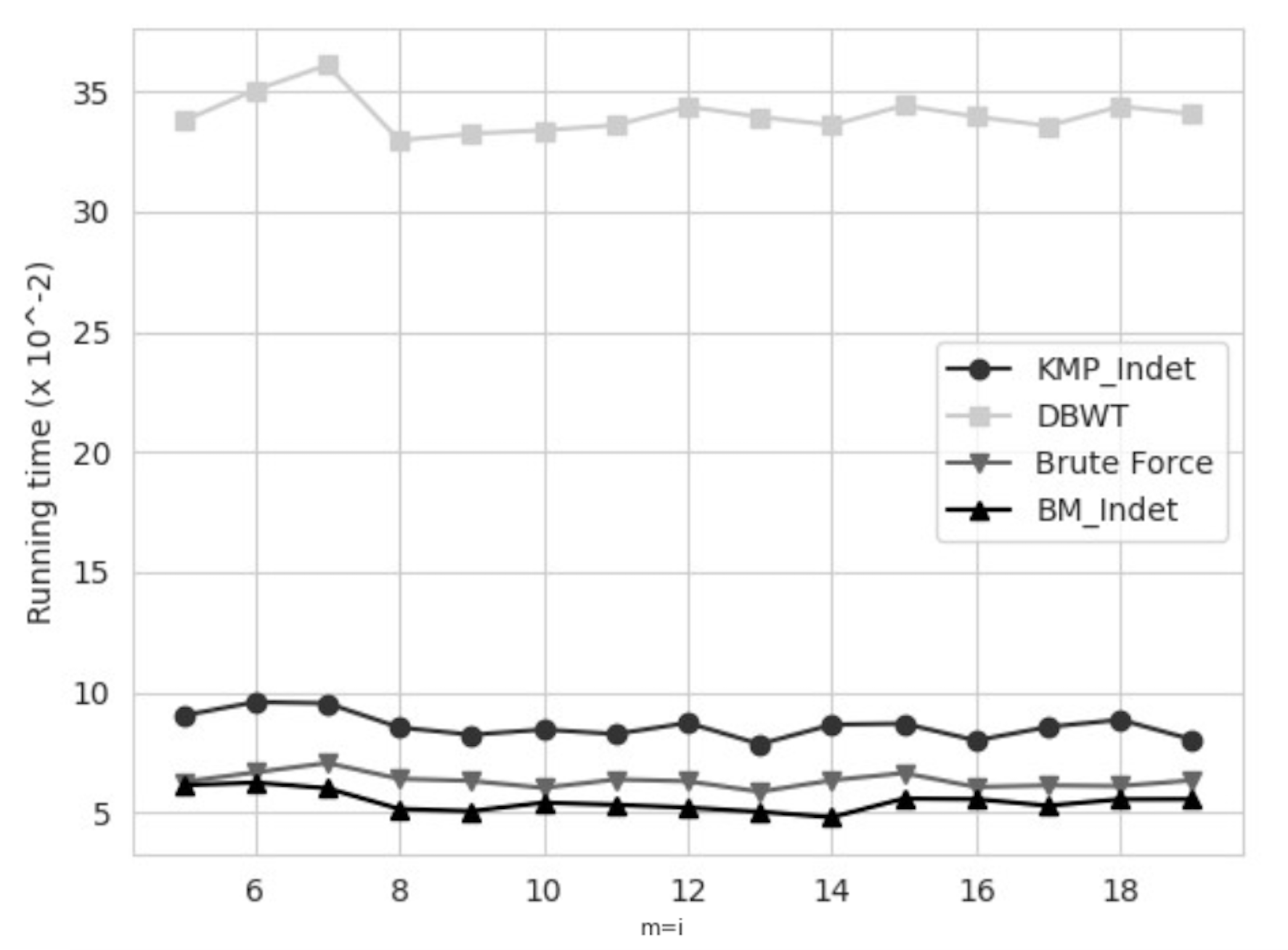}
            \end{minipage}
            \begin{minipage}{.45\textwidth}
              \centering
               \includegraphics[scale=.38]{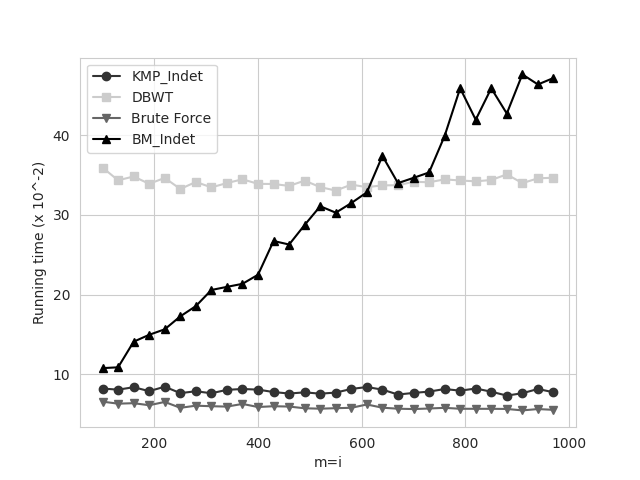}
            \end{minipage}
           
          \caption{For $\sigma=4, n = 10^6, k_1=0$, where $m = i, k_2  = i/5$, $1 \leq i \leq 19$ (for graph on the left) and $m = i, k_2  = i/10$, where $100 \leq i \leq 1000$ (for graph on the right).}
        \label{fig:dbwt1}
         \end{figure}

    \begin{figure}[h!]
             \centering
            \begin{minipage}{.45\textwidth}
                \centering
               \includegraphics[scale=.37]{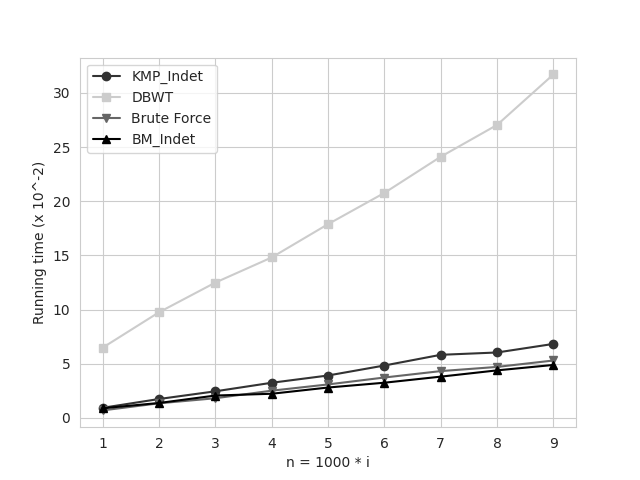}
            \end{minipage}
            \begin{minipage}{.45\textwidth}
              \centering
               \includegraphics[scale=.37]{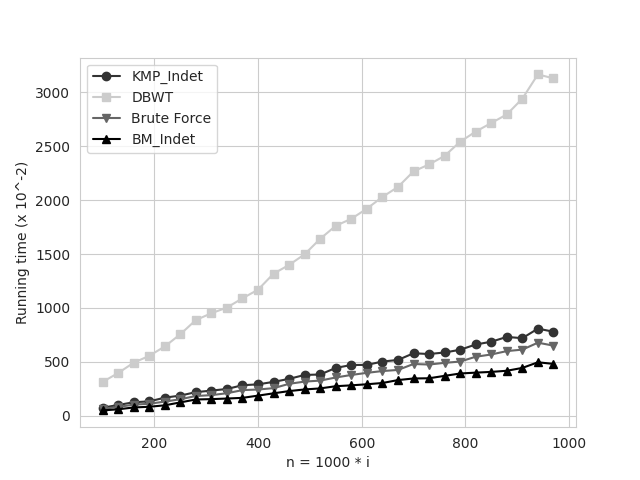}
            \end{minipage}
             \caption{For $\sigma = 4, n=1000i$, where $1\leq i \leq 10$ (for graph on the left) and $100\leq i \leq 1000$ (for graph on the right), $m=20, k_1=0, k_2=2$.}
            \label{fig:dbwt2}
             \end{figure}

From the graphs shown in Figures~\ref{fig:dbwt1} and \ref{fig:dbwt2}, we observe that for fixed length texts ($n=10^6$), short texts ($n=1000i$, where $1\leq i\leq10$) and long texts ($n=1000i$, where $100\leq i\leq1000$), when the pattern length is small ($m\leq 20$), \textsc{BM\_Indet} performs better than \textsc{KMP\_Indet} and \textsc{BF}, and performs significantly better than \textsc{DBWT}. However, as the length and the number of indeterminate letters in the pattern increases, \textsc{BM\_Indet} performs worse than all the three algorithms, including \textsc{DBWT}, while \textsc{KMP\_Indet} and \textsc{BF} continue to perform better than \textsc{DBWT}. This is because, with increasing pattern size, the size of the bad character table also increases and thus requires more time to compute it. Furthermore, as the number of indeterminate letters increase in the pattern, the likelihood of \textsc{BM\_Indet} requiring to compute the prefix array increases. Moreover, irrespective of the length of the matched substring in $\s{y}$, $|\s{q}'|$ computed by \textsc{BM\_Indet} is always greater than or equal to $|\s{q}'|$ (and hence the prefix array) computed by \textsc{KMP\_Indet}. Hence, due to these factors \textsc{BM\_Indet} requires more time to compute the prefix array at each iteration than \textsc{KMP\_Indet}. Therefore, \textsc{BM\_Indet} performs worse with increasing pattern lengths and an increasing number of indeterminate letters.

\subsection{Comparison of \textsc{KMP\_Indet}, \textsc{BM\_Indet} and \textsc{BF} Algorithms}\label{ssec:comp2}

In all the graphs generated in this section, the text length $n$ is a multiple of $i$ (where $1 \leq i \leq 10$ for short texts, and $100\leq i \leq 1000$ for long texts), the pattern length $m = 40i$, and the numbers of indeterminate letters in text and pattern are $k_1 = 0.06i, k_2=4i$, respectively. 

        \begin{figure}[h!]
         \centering
        \begin{minipage}{.45\textwidth}
            \centering
            \includegraphics[scale=.35]{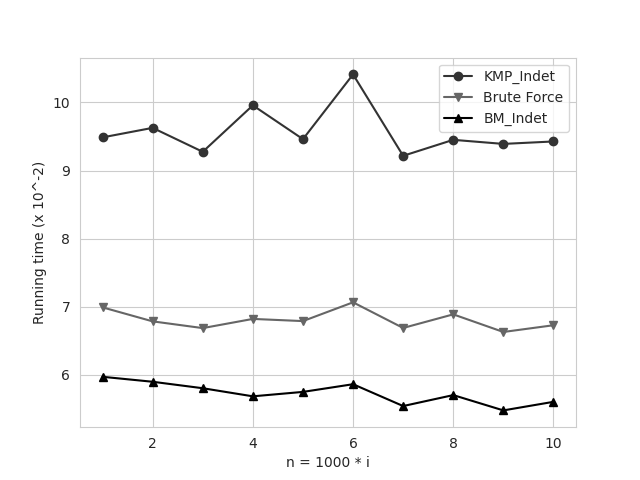}
        \end{minipage}
        \begin{minipage}{.45\textwidth}
          \centering
            \includegraphics[scale=.4]{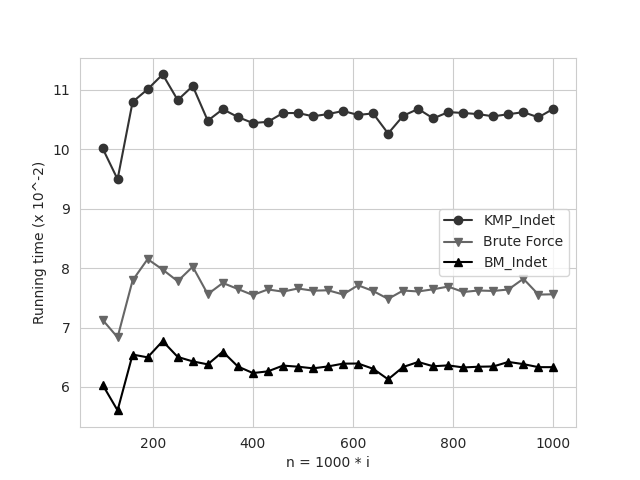}
        \end{minipage}
         \caption{For $\sigma = 2, n=1000i$, where $1\leq i \leq 10$ (for graph on the left) and $100\leq i \leq 1000$ (for graph on the right), $m=40i, k_1=0.06n, k_2=4i$.}
        \label{fig:case1}
         \end{figure}

\begin{figure}[h!]
 \centering
\begin{minipage}{.45\textwidth}
    \centering
    \includegraphics[scale=.35]{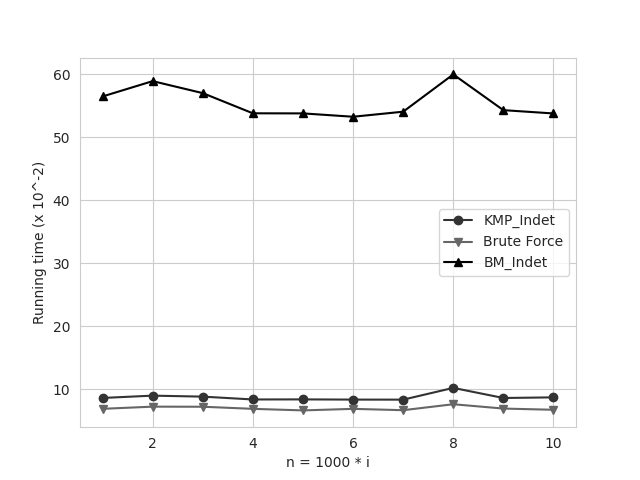}
\end{minipage}
\begin{minipage}{.45\textwidth}
  \centering
    \includegraphics[scale=.4]{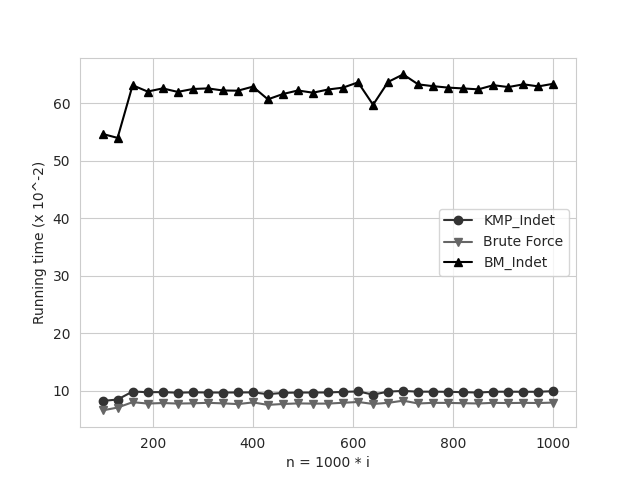}
\end{minipage}
\caption{For $\sigma = 4, n=1000i$, where $1\leq i \leq 10$ (for graph on the left) and $100\leq i \leq 1000$ (for graph on the right), $m=40i, k_1=0.06n, k_2=4i$.}
\label{fig:case2}
 \end{figure}

\begin{figure}[!h]
 \centering
\begin{minipage}{.45\textwidth}
    \centering
    \includegraphics[scale=.35]{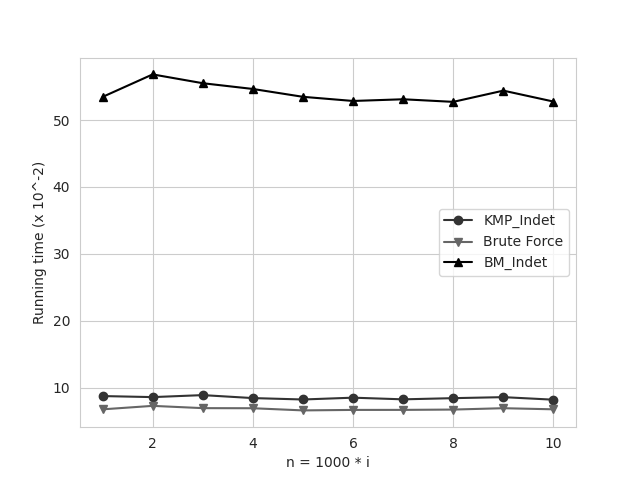}
\end{minipage}
\begin{minipage}{.45\textwidth}
  \centering
    \includegraphics[scale=.4]{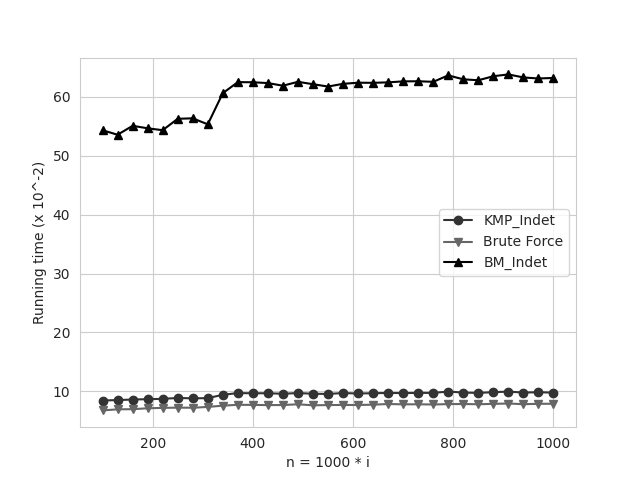}
\end{minipage}
\caption{For $\sigma = 9, n=1000i$, where $1\leq i \leq 10$ (for graph on the left) and $100\leq i \leq 1000$ (for graph on the right), $m=40i, k_1=0.06n, k_2=4i$.}
\label{fig:case3}
 \end{figure}
From the experiments conducted we observe that for indeterminate strings on a binary alphabet, \textsc{BM\_Indet} is significantly faster than \textsc{KMP\_Indet} and \textsc{BF}: see Figure~\ref{fig:case1}. This is because, the bad character table is small in size and therefore can be computed efficiently. On the other hand, with larger alphabet sizes ($3\leq \sigma \leq 9$), \textsc{KMP\_Indet} and \textsc{BF} perform significantly better than \textsc{BM\_Indet} see Figures~\ref{fig:case2} and~\ref{fig:case3}. The poor performance of \textsc{BM\_Indet} is again due to the construction of large bad character tables and prefix arrays resulting in increased computational time.

\section{Conclusion}\label{sec:conclusion}
We have described simple procedures, based on the KMP and BM algorithms, to do pattern-matching on indeterminate strings, both implemented so
as to require very little additional storage.
We observe that \textsc{KMP\_Indet} performs similar to the brute force (\textsc{BF}) algorithm on random strings over alphabet sizes $\geq 4$. \textsc{BM\_Indet}, on the other hand, seems to perform better than \textsc{KMP\_Indet} and \textsc{BF} on binary strings and on smaller patterns ($m \le 20$) over alphabet sizes of $3$ and $4$, but worse on strings over alphabet sizes $>4$. 
We also compared the performance of \textsc{KMP\_Indet} and \textsc{BM\_Indet} with the only competing algorithm (\textsc{DBWT})~\cite{DGGL19} with a currently available implementation; we discover that \textsc{KMP\_indet} executes an order of magnitude faster than \textsc{DBWT} in (almost) all the cases tested. Moreoever, the \textsc{BM\_Indet} algorithm performs better than \textsc{DBWT} on smaller patterns and on those with fewer indeterminate letters. The main reason for this advantage is the avoidance of elaborate data structures in both the \textsc{KMP\_Indet} and \textsc{BM\_Indet} algorithms. We further discover that (surprisingly) \textsc{BF} is generally the fastest of the three! The consequences of these observations need further investigation; in particular, it would be of interest to study the performance of these algorithms on real data; in particular, indeterminate strings that arise in practice (especially DNA strings on $\sigma = 4$).

Based on our experiments we observed that the primary reason for the poor performance of \textsc{BM\_Indet} was due to the increasing resources needed to construct the bad character table with increasing alphabet sizes and pattern lengths. We believe that optimizing the bad character table construction will significantly improve \textsc{BM\_indet} algorithm's performance. Moreover, it is also of interest to implement and test indeterminate versions of the several variants of the  Boyer-Moore algorithm (BM-Horspool, BM-Sunday, BM-Galil, Turbo-BM): see \cite[Ch.\ 8]{S03} and
\begin{center}
\url{https://www-igm.univ-mlv.fr/~lecroq/string/}
\end{center}
to see if the \textsc{BM\_indet} algorithm's performance further improves with these variations.

\section*{Acknowledgements}

The first and third authors were funded by the Faculty of Engineering and the Department of Computing and Software, McMaster University. The fourth author was supported by the Natural Sciences \& Engineering Research Council of Canada (NSERC) [Grant No. 10536797]. 

\bibliographystyle{elsarticle-num}
\bibliography{references}
\end{document}